# Unconventional Flatband Line States in Photonic Lieb Lattices

Shiqi Xia[1], Ajith Ramachandran[2], Shiqiang Xia[1], Denghui Li[1], Xiuying Liu[1], Liqin Tang[1],

Yi Hu[1], Daohong Song[1,3], Jingjun Xu[1,3], Daniel Leykam[2], Sergej Flach[2], and Zhigang Chen[1,3,4]

[1]*The MOE Key Laboratory of Weak-Light Nonlinear Photonics, TEDA Applied Physics Institute and School of Physics, Nankai University, Tianjin 300457, China*
[2]*Center for Theoretical Physics of Complex Systems, Institute for Basic Science (IBS), Daejeon 34126, Republic of Korea*
[3]*Collaborative Innovation Center of Extreme Optics, Shanxi University, Taiyuan, Shanxi 030006, People's Republic of China*
[4]*Department of Physics and Astronomy, San Francisco State University, San Francisco, California 94132, USA*
songdaohong@nankai.edu.cn, dleykam@ibs.re.kr, sflach@ibs.re.kr, zgchen@nankai.edu.cn,

**Abstract:** Flatband systems typically host "compact localized states" (CLS) due to destructive interference and macroscopic degeneracy of Bloch wave functions associated with a dispersionless energy band. Using a photonic Lieb lattice (LL), we show that conventional localized flatband states are inherently incomplete, with the missing modes manifested as extended *line states* which form non-contractible loops winding around the entire lattice. Experimentally, we develop a continuous-wave laser writing technique to establish a finite-sized photonic LL with specially-tailored boundaries, thereby directly observe the unusually extended flatband line states. Such unconventional line states cannot be expressed as a linear combination of the previously observed CLS but rather arise from the nontrivial real-space topology. The robustness of the line states to imperfect excitation conditions is discussed, and their potential applications are illustrated.

Flatband systems, first proposed for the study of ferromagnetic ground states in multiband Hubbard models, have proven to be conceptually effective and important in condensed matter physics [1-3]. They are characterized by a band structure with one band being completely flat, signaling macroscopic degeneracy. One can construct CLS which remain intact during evolution due to destructive interference. Over the years, a variety of approaches have been developed to design and characterize different flatband systems [4-8], with lattice geometries ranging from sawtooth, stub, diamond, dice, kagome, to Lieb and perovskite lattices in general [7-12]. This is largely due to the flatband systems providing a platform for probing various fundamental phenomena that have intrigued scientists for decades, including Anderson localization [6, 13, 14], nontrivial topological phases and quantum Hall states [15-19], and flatband superfluidity [20, 21].

The Lieb lattice (LL) – a face-centered square depleted lattice [Fig. 1(a)] – is geometrically different from other two-dimensional lattices such as square and honeycomb lattices. This peculiar system possesses a single conical intersection point in its Brillouin zone (BZ), where the flatband is sandwiched between two conical Bloch bands [Fig. 1(b)]. The flatband in the LL is protected by a chiral symmetry, and its intersection with the dispersive bands is protected by real-space topology [12, 22, 23]. Recently, LLs have been realized in several different settings, including Bose-Einstein condensates [4, 24], surface state electrons [25, 26], exciton-polaritons in micropillars [27], and waveguide arrays in photonic structures [28-32]. However, so far most of previous experimental studies have focused on the demonstration of the LL structures and their associated CLS, overlooking unusual features that arise in infinitely extended lattices [Fig. 1(c)] or finite (truncated) lattices with different cutting boundaries [Figs. 1(d, e)].

In this Letter, we demonstrate the CLS previously investigated in the LL are linearly dependent, so they do not form a complete basis for the flatband. The missing states completing the flatband basis are unconventional line states, which *cannot* be formed by linear superpositions of the conventional CLS [23]. Such line states wrap around the lattice under periodic boundary conditions, forming noncontractible loops. However, with appropriate termination of the lattice, these line states can be observed even in a finite system. Experimentally, we develop a simple yet effective writing technique with a weak continue-wave laser (rather than a high-power femtosecond laser) to establish finite-sized photonic LL with desired boundaries. More importantly, we demonstrate the existence of the line states (and their variation in other shapes) in the truncated LL from direct measurements of both real-space intensity and momentum-space spectrum. The robustness of these unconventional line states to imperfect excitation conditions is also discussed, along with specific examples of their potential applications.

Propagation of a light beam along the LL composed by evanescently coupled identical optical waveguides [sketched in Fig. 1(a)] is described by the paraxial wave equation [32, 33]: $\frac{\partial}{\partial z}\psi(x,y,z) = \left(\frac{i}{2k}\nabla_\perp^2 + \frac{ik}{n}\Delta n(x,y)\right)\psi(x,y,z)$, where $\psi(x,y,z)$ is the envelope of the electric field, $k = \lambda n/2\pi$ is the wave number in the medium, $\lambda$ is the wavelength, $n$ is the bulk refractive index, $\Delta n(x,y)$ is the optically-induced refractive index lattice, and $\nabla_\perp^2 = \partial_x^2 + \partial_y^2$ is the transverse Laplacian. Under the tight-binding approximation, this model reduces to the well-known linear discrete Schrödinger equation [28-31, 33]. For the LL, eigenvalue equations can be explicitly written for wave amplitudes in the unit cell denoted by (n, m) as,

$$\mathbf{E} A_{n,m} = B_{n,m} + B_{n,m-1} + C_{n,m} + C_{n-1,m},$$
$$\mathbf{E} B_{n,m} = A_{n,m} + A_{n,m+1}, \quad \mathbf{E} C_{n,m} = A_{n,m} + A_{n+1,m}. \quad (1)$$

A plane wave ansatz for the eigenvectors with momenta ($k_x$, $k_y$) yields the band structure: $\mathbf{E}_{FB} = 0$, $\mathbf{E}(k_x, k_y) = \{\pm(2(2+\cos(k_x)+\cos(k_y))^{1/2}\}$. A zero energy flatband touches two linearly dispersive bands intersecting at the conical intersection point as shown in Fig. 1(b). Normally, modes in the continuum are not localized, except for those generated under special conditions as "bound states in the continuum" [34-36]. The flatband supports inherently degenerated CLS that have nonzero amplitude only at four lattice sites as shown in Fig. 1(a), where all *A* sites (the minority sites) have zero amplitude but *B* and *C* sites (the majority sites) have nonzero amplitude with opposite phase. Such a solution satisfies Eq. (1) at **E**=0. These states are not only localized but also compact, residing in only three unit cells. Other localized states (such as necklace-shaped flatband states can be established by linear combinations of the CLS, as demonstrated in previous experiments [30-32]. However, in certain lattices where the flatband intersects other dispersive bands, the set of CLS can become linearly dependent under periodic boundary conditions, i.e. an appropriate superposition of all the CLS vanishes, and consequently they do not form a complete basis for the flatband [23]. The missing states which complete the flatband basis are the *line states*, which are extended across the entire lattice. Such line states also satisfy conditions for destructive interference, and their associated wave functions represent exact eigenstates at the flatband energy **E**=0. A line-state solution of Eq. (1) elongated in the *y*-direction is given by $B_{n,m} = -B_{n,m+1}$ for a given n, with all other amplitudes in Eq. (1) vanishing. Likewise, a line state elongate in the x-direction is given by $C_{n,m} = -C_{n+1,m}$ for a given m, and all other amplitudes vanish. These states cannot be expressed as a linear combination of the CLS shown in Fig. 1(a). For an infinite lattice (or a lattice with periodic boundary conditions folds into a 2D torus), there are only two distinct line states as illustrated in Fig. 1(c), and all other line states can be obtained by discrete translation or combination. For an experimentally feasible finite-sized LL, the termination of the lattice boundaries determines whether a particular line state can exist. For the "flat" edges shown in Fig. 1(d), only the necklace-shaped state is allowed, but this state can be expressed as a linear superposition of the CLS. Linearly independent line states do not exist and the CLS form a complete basis under this edge condition. On the other hand, by cutting the lattice to form "bearded" edges [37], the line state terminates smoothly at the boundaries [Fig. 1(e)]. In this latter case, the line state can be efficiently kept as a single line, but it cannot be expressed as a superposition of the CLS. The *k*-space spectra of the CLS and the line state are measured and displayed in Figs. 1(f, g), where the white dashed square marks the edge of the first BZ. The spectrum of the CLS is distributed around the four sides of the 1$^{st}$ BZ, whereas that for the line state mainly along the two zone edges. In Figs. 1(h, i), a band structure calculation using the paraxial equation is shown for a semi-infinite lattice, illustrating the existence/absence of edge states depending on the lattice termination. A detailed explanation of boundary-dependent superposition of the CLS in finite LL is provided in the Supplemental Material, emphasizing the linear dependence (independence) of the CLS for the "bearded" ("flat") edges [12].

To observe the extended flatband states, we introduce a simple cw-laser writing technique to establish the finite-sized photonic LL with desired boundaries. The technique relies on site-to-site induction of waveguides in a nonlinear photorefractive (SBN) crystal with noninstantaneous nonlinear response. Different from the femtosecond laser writing waveguides in glass [28-31], the Lieb lattice optically induced in the crystal can be readily reconfigurable. Moreover, our writing technique can generate lattices with virtually arbitrary lattice edges, which is not possible with the previous technique based on multi-beam interference [32]. Figure 2 shows a schematic of the experimental setup. A cw laser beam with a wavelength of 532nm and an output power of only 50mW is used to illuminate a phase-only spatial light modulator (SLM), which creates a Gaussian beam with reconfigurable input positions at the crystal input facet. The 4F system guarantees a quasi-non-diffracting zone of the writing beam as it propagates through the 10mm-long crystal. Because of the noninstantaneous self-focusing nonlinearity, the writing beam induces waveguides one-by-one through the crystal and all waveguides remain intact within the experimental data acquisition period. The bottom panels in Fig. 2 illustrates the writing principle and a typical triangular lattice written in the crystal. (A non-diffracting Bessel beam can also be used to write the waveguides, but the side lobes of the Bessel beam can affect the lattice structure [38]).

After this proof of principle, we employ the technique to write photonic LL with desired lattice edges. Figure 3(a) shows such a laser-written LL with "bearded" edges. The lattice spacing is about 32 μm. With controlled writing beam intensity and exposure time, the waveguide coupling occurs mainly between nearest neighbors, satisfying the tight-binding model. Indeed, after propagating through the 10mm crystal, a single-site excitation leads to discrete diffraction and coupling mainly to the nearest waveguides [see inset in Fig. 3(a)]. To observe the extended flatband line states displayed in Fig. 1(e), the input probe beam is shaped into a line pattern (a broken and stretched "necklace"), with its phase modulated by the SLM so that adjacent "pearls" have opposite phase [Fig. 3(b1)]. Without the lattice, the line shape cannot be preserved after free propagation due to interference between "pearls" and diffraction of each "pearl" [Figs. 3(b2)]. In contrast, when such a line beam is launched into the LL [see in Fig. 3(a) for its input position], its overall intensity pattern is well-maintained [Fig. 3(b3)]. Furthermore, each "pearl" remains localized and out of phase with its neighbors as verified from the interferogram [Fig. 3(b4)], in agreement with the predictions. For direct comparison, corresponding results for an in-phase line beam (when all "pearls" are made with equal phase) are presented in Figs. 3(c1-c3). In this latter case, the line becomes deteriorated, as energy couples to zero-amplitude lattice sites [Fig. 3(c3)], although the in-phase feature is still preserved [Fig. 3(c4)]. Moreover, dramatic difference is observed in the $k$-space spectrum between the out-of-phase and in-phase line beams [Figs. 3(b5, c5)]: The spectrum of out-of-phase line is distributed along the BZ edges, whereas that of in-phase one goes to the center of the 1$^{st}$ BZ, although tunneling to higher BZs is present in both cases because of coupling to dispersive bands in the finite LL. Because of the short propagation distance limited by crystal length in experiment, we perform numerical simulation to further corroborate the experimental observations. Results with parameters similar to our experiments but for a much longer propagation distance (40 mm) are presented in [Figs. 3(d, e)]. One can see clearly the difference between the out-of-phase beam (remains intact as unconventional line states) and in-phase beam (becomes strongly distorted). Following the similar approach we also generated a photonic LL with "flat" edges and demonstrated the necklace-shaped flatband states illustrated in Fig. 1(d). These results and relevant discussions are provided in the Supplementary Material.

To illustrate the robustness of the line states to imperfect excitation conditions in the LL, we perform additional experiments and numerical simulations based on the tight-binding model. Typical results are presented in Fig. 4. A "shorter" line (not winding around all the way to lattice edges) tends to extend and couple its energy to the two "bearded" edges to form the line state [Figs. 4(a)]., but it cannot preserve in the LL with "flat" edges as its energy couples more to the direction perpendicular to the line rather than along the line [Fig. 4(b)]. These results show that the line states are boundary-dependent eigenstates, quite different from the conventional boundary-independent CLS, as discussed further in the Supplementary Material. Fully extended lines with "imperfect" initial phase or amplitude can still evolve into "self-healing" line states [Figs. 4(c, d)]. Of course, if the input conditions deviate too much from that of the line states, the dispersive band modes are excited mostly which leads to strong discrete diffraction in the lattice.

Before closing, we would like to elaborate on potential applications of these unconventional flatband line states. Conventional CLS can be useful for image transmission in the LL, but only limited to certain necklace-like shape as superposition of "ring modes" [32]. By combining the features of the CLS and the new line states, it is possible to realize large-scale image transmission with virtually any patterns extended to lattice boundaries [39]. To illustrate this, we transmit three different letters ("PRL") based on the combination of the line states and the CLS in the "bearded" LL, as shown in Fig. 5. All three letters consist of out-of-phase points, terminating on the "bearded" edges [Fig. 5(a)]. The shape of the letters preserves after propagation through the lattice, as observed in experiment [Fig. 5(b)]. Numerical simulation to a longer propagation distance (40mm) leads to dramatic difference for direct comparison between the out-of-phase and in-phase letters [Figs. 5(c, d)]. The measured output $k$-space spectra for all three letters are qualitatively similar (only the spectrum of letter L is shown in the inserts), distributing mainly along the $1^{st}$ BZ edges for the out-of-phase letters (i.e., based on the flatband states), but splitting into the center and higher zones for the in-phase letters where modes of dispersive bands are excited. Another example is related to tunable conductivity. As shown in the band structure for a semi-infinite LL [Fig. 1(f, g)], nanoribbons can be made conducting or insulating. Moreover, the state in Fig. 1(d) could be destroyed by a vertical potential gradient (electric field), while the state in Fig. 1(e) would be completely unaffected. Therefore, nanoribbons formed from artificial LL can be either metallic

(gapless) or semiconducting (gapped), depending on the edge termination. This generalizes the behavior of graphene nanoribbons. By tuning the potential of the edge sites, one may switch between metallic and semiconducting regimes, or "pump" modes between the two bulk bands.

In conclusion, we have proposed and demonstrated unconventional flatband line states in a photonic LL "fabricated" with desired boundaries. Our experimental results are further corroborated by numerical simulations based on the paraxial wave equation, while the robustness of the line states to imperfect excitation conditions is studied with the coupled mode theory. Our work demonstrates that reconfigurable platform of designer photonic lattices could bring about numerous opportunities for both fundamental studies and potential applications. For example, the unconventionally extended flatband states may prove relevant to similar phenomena in other flatband systems in condensed matter physics. Moreover, our technique of writing special LL in nonlinear crystals could be useful to study a number of intriguing fundamental phenomena such as edge states and topological phases, singularity character and flatband topology, and nonlinear topological solitons [27, 40-43].

Acknowledgement: This work is supported by the National Key R&D Program of China (2017YFA0303800), the Chinese National Science Foundation (91750204 and 11674180), by PCSIRT（IRT_13R29）and 111 Project (No. B07013) in China, and the Institute for Basic Science in Korea (IBS-R024-D1 and IBS-R024-Y1).

**Reference:**

[1] E. H. Lieb, Phys. Rev. Lett. **62**, 1201 (1989).

[2] A. Mielke, J. Phys. A **24**, L74 (1991); **24**, 3311 (1991); **25**, 4335 (1992).

[3] H. Tasaki, Eur. Phys. J. B **64**, 365 (2008).

[4] F. Baboux et al., Phys. Rev. Lett. **116**, 066402 (2016).

[5] W. Maimaiti et al., Phys. Rev. B **95**, 115135 (2017)

[6] M. Goda et al, Phys. Rev. Lett. **96**, 126401 (2006).

[7] M. Hyrkäs et al, Phys. Rev. A **87**, 023614 (2013).

[8] S. D. Huber and E. Altman, Phys. Rev. B **82**, 184502 (2010).

[9] V. Apaja et al, Phys. Rev. A **82**, 041402 (2010).

[10] D. Green et al, Phys. Rev. B **82**, 075104 (2010).

[11] C. Weeks et al, Phys. Rev. B **82**, 085310 (2010).

[12] D. L. Bergman et al, Phys. Rev. B **78**, 125104 (2008).

[13] D. Leykam et al, Phys. Rev. B **88**, 224203 (2013).

[12] J. T. Chalker, et al, Phys. Rev. B **82**, 104209 (2010).

[15] E. J. Bergholtz and Z. Liu, Int. J. Mod. Phys. B **27**, 1330017 (2013)

[16] S. A. Parameswaran et al, Physique **14**, 816 (2013).

[17] T. Neupert et al, Phys. Rev. Lett. **106**, 236804 (2011).

[18] K. Sun et al, Phys. Rev. Lett. **106**, 236803 (2011).

[19] E. Tang et al, Phys. Rev. Lett. **106**, 236802 (2011).

[20] S. Peotta and P. Törmä, Nat. Commun. **6**, 8944 (2015).

[21] A. Julku et al, Phys. Rev. Lett. **117**, 045303 (2016).

[22] R. Shen et al, Phys. Rev. B **81**, 041410(R) (2010)

[23] A. Ramachandran et al, Phys. Rev. B **96**, 161104(R) (2017).

[24] S. Taie et al, Sci Adv. **1**, 1500854 (2015).

[25] W.-X. Qiu et al. Phys. Rev. B **94**, 241409 (2016).

[26] M. R. et al, Nat. Phys. **13**, 672 (2017).

[27] S. Klembt et al., Appl. Phys. Lett. **111**, 231102 (2017);
    C. E. Whittaker et al., Phys. Rev. Lett. **120**, 097401 (2018).

[28] F. Diebel et al, Phys. Rev. Lett. **116**, 183902 (2016).

[29] D. Guzmán-Silva et al, New J. Phys. **16**, 063061 (2014).

[30] R. A. Vicencio et al, Phys. Rev. Lett. **114**, 245503 (2015).

[31] S. Mukherjee et al, Phys. Rev. Lett. **114**, 245504 (2015).

[32] S. Xia et al, Opt. Lett. **41**, 1435 (2016).

[33] F. Lederer et al., *Phys. Rep.* **463,** 1 (2008).

[34] C. W. Hsu et al, Nat. Rev. Mater. **1**, 16048 (2016).

[35] D. C. Marinica et al, Phys. Rev. Lett. **100**, 183902 (2008).

[36] Y. Plotnik et al, Phys. Rev. Lett. **107**, 183901(2011).

[37] Y. Plotnik et al, Nat. Mater. **13**, 57 (2014).

[38] F. Diebel, et al., Appl. Phys. Lett. **104**, 191101 (2014).

[39] R. A Vicencio, C Mejía-Cortés, Journal of Optics **16**, 015706 (2013).

[40] C. Li et al, Phys. Rev. B **97**, 081103(R) (2018).

[41] R. Chen et al, Phys. Rev. B **96**, 205304 (2017).

[42] J. Noh et al, Phys. Rev. Lett. **120**, 063902 (2018).

[43] J. Rhim and B. Yang, https://arxiv.org/abs/1808.05926.

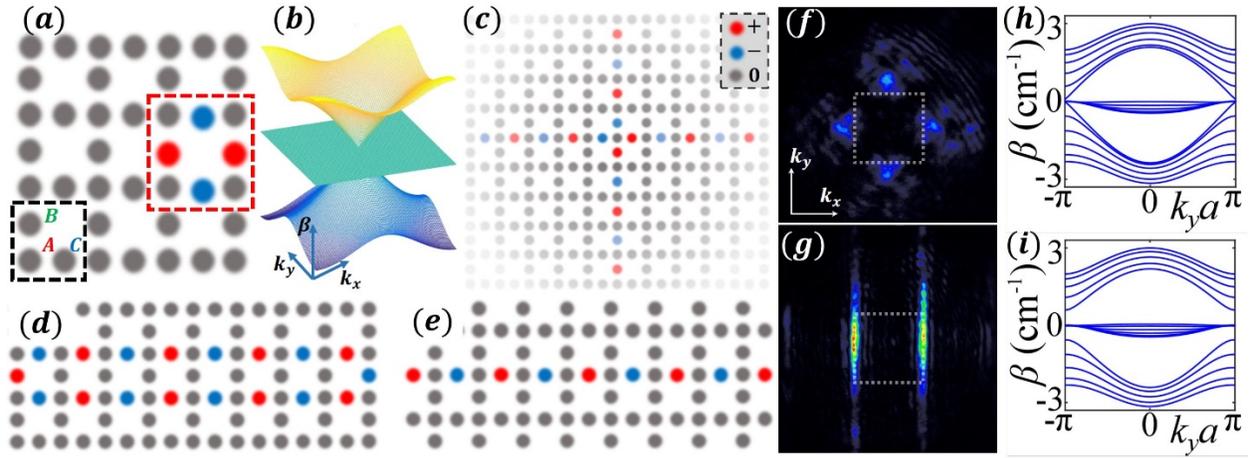

FIG. 1: Flatband states in a photonic LL. (a) Illustration of lattice structure consisting of three sublattices ($A,B,C$) shown in dark-dashed square, with a fundamental flatband mode (the CLS) shown in red-dashed square. Sites with zero amplitudes are denoted by gray dots, and those of nonzero amplitudes with opposite phase by red and blue dots. (b) Band structure of the lattice in the 1st BZ. (c) Unconventional line states (oriented horizontally and vertically) in an *infinite* LL. (d, e) Necklace-shaped flatband states (unconventional line states) in a *finite* LL with "flat" ("bearded") edges. (f, g) Measured $k$-space spectrum of (f) the CLS and (g) the line state. (h, i) Calculated band structure for a semi-infinite LL with (h) "flat" and (i) "bearded" edges.

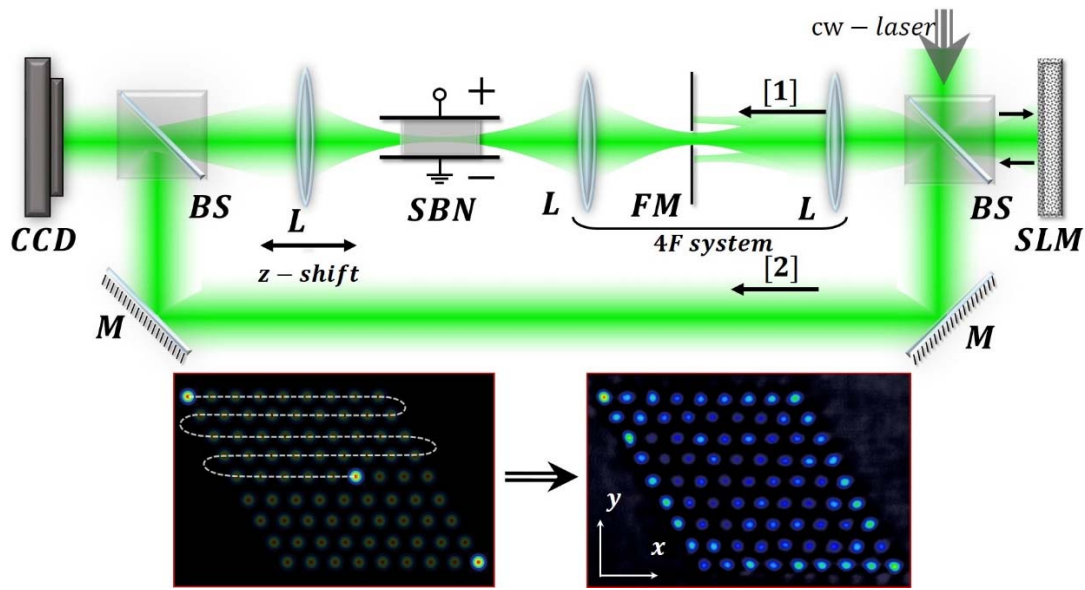

FIG. 2 (color online). Experimental setup for point-to-point writing of photonic lattices with a cw-laser in a nonlinear crystal. SLM: spatial light modulator; BS: beam splitter; FM: Fourier mask; SBN: strontium barium niobate. Path 1 is for both writing and probe beams, and path 2 is for reference beam for output interference measurement. Bottom left: illustration of the sequential positions of the writing beam. Bottom right: a typical photonic lattice written in the crystal as probed by a broad (quasi-plane-wave) beam.

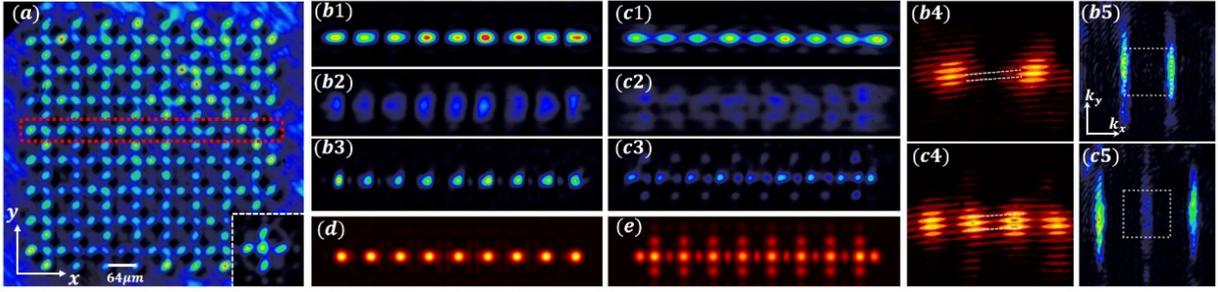

Fig. 3: Demonstration of unconventional line state in a photonic LL with "bearded" edges. (a) LL established with cw-laser writing technique. Red dashed square marks the input position of the line-shaped probe beam (the inset shows discrete diffraction of single-site excitation). (b1-b3) Transverse intensity pattern of out-of-phase beam at (b1) input, (b2) output without LL, and (b3) output through LL as a propagation-invariant eigenstate. (b4) Separate zoom-in interferogram of output pattern with a reference beam showing adjacent "pearls" in the line beam are out of phase (white dashed lines added for visualization). (b5) Measured $k$-space spectrum of (b3) with dashed square marking the 1st BZ. (c1-c5) Same as in (b1-b5) except that all sites are now in-phase. (d, e) Simulation results showing out-of-phase line state remains intact but in-phase one deforms strongly after propagating a distance of 4 cm through the lattice.

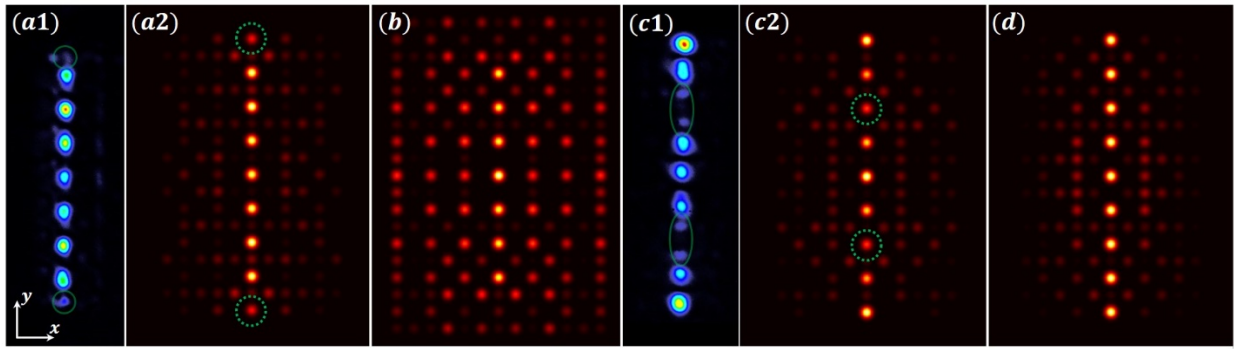

Fig. 4: Construction and destruction of the line state under imperfect excitation conditions. A "shorter" line not extending to the LL boundary tends to evolve into a line state with the "bearded" edges (a1, from experiment; a2, from simulation), but deforms strongly with the "flat" edges (b). (c, d) Restoring of line states from a fully extended line beam but with "imperfect" amplitude, e.g., with two broken points (c1, from experiment; c2, from simulation) or "imperfect" phase (e.g., $3\pi/4$) between adjacent sites (d). The green dashed circles mark the initial empty sites at input, but with coupled intensity at output. The restoration of the line state is not complete in (a1 and c1) due to limited propagation distance (1 cm) in experiment.

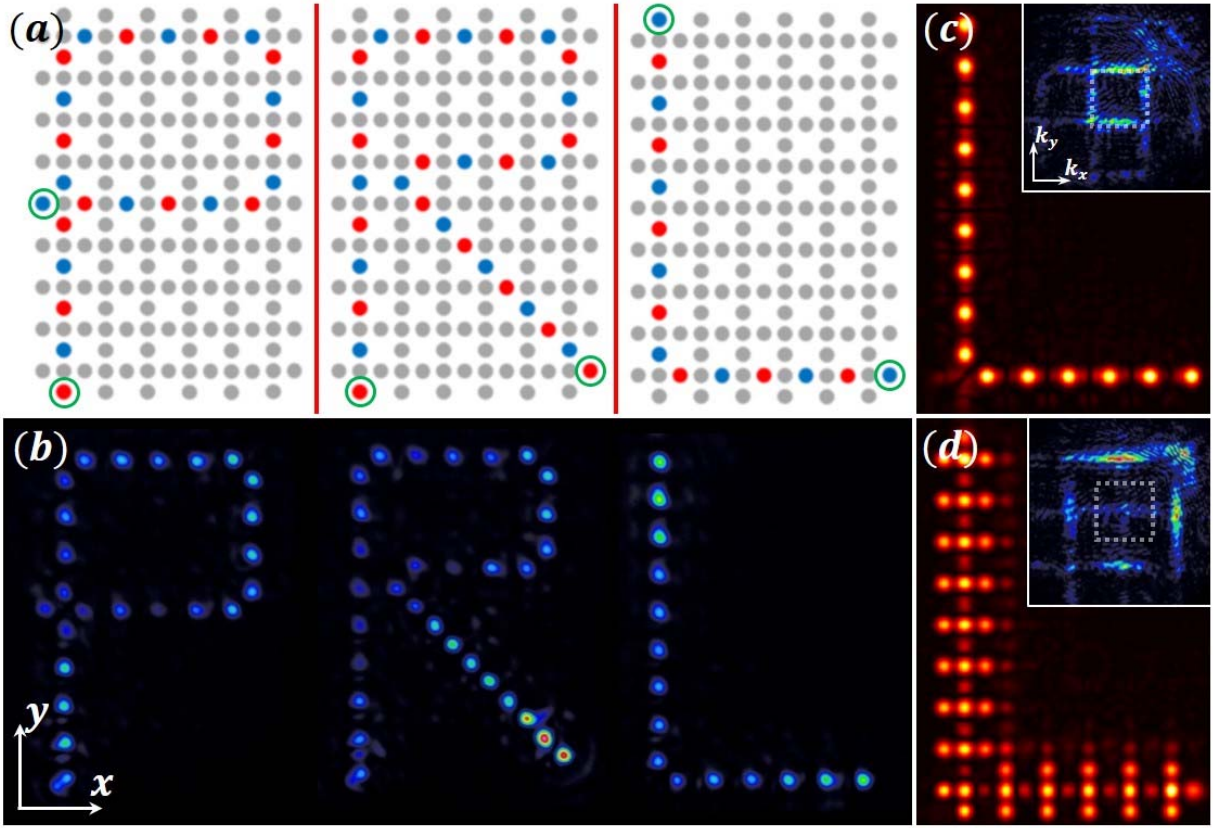

Fig. 5: Demonstration of text transmission based on flatband line states. (a) Schematic of the input letters "PRL" in a "bearded" Lieb lattice, with the ending sites marked by green circles. (b) Experimental results of output letters after propagating through the lattice. (c, d) Numerical simulation to a propagation distance (4 cm) for direct comparison between (c) out-of-phase and (d) in-phase letter "L", where the inserts show the measured Fourier spectrum in experiment.